\documentclass[usenatbib]{mn2e}

\usepackage{graphicx}
\usepackage{amsmath}
\usepackage{lscape}
\usepackage{afterpage}
\usepackage{soul}
\usepackage{color}

\title[A bright ULX in NGC 5907]{A bright ultraluminous X-ray source in NGC 5907}

\author[A.\,D. Sutton et al.]
{
Andrew D. Sutton$^1$\thanks{Email: andrew.sutton@durham.ac.uk}, Timothy P. Roberts$^1$, Jeanette C. Gladstone$^2$, Sean A. Farrell$^{3,4}$, \and Emma Reilly$^{3,5}$, Michael R. Goad$^3$, Neil Gehrels$^6$\\
 \\
$^1$Department of Physics, University of Durham, South Road, Durham, DH1 3LE, UK\\
$^2$Department of Physics, University of Alberta, Edmonton, Alberta, T6G 2G7, Canada\\
$^3$Department of Physics and Astronomy, University of Leicester, University Road, Leicester, LE1 7RH, UK\\
$^4$Sydney Institute for Astronomy, School of Physics, The University of Sydney, NSW 2006, Australia\\
$^5$Astrophysics Research Centre, School of Mathematics and Physics, Queens University Belfast, Belfast, BT7 1NN, UK\\
$^6$NASA Goddard Space Flight Center, Greenbelt, MD 20771, USA\\
}

\pagerange{\pageref{firstpage}--\pageref{lastpage}} \pubyear{2011}

\newcommand{\chan}{{\it Chandra~\/}}

\def\Msun{\hbox{$\rm M_{\odot}$}}

\def\H0{{\rm ~km~s^{-1}~Mpc^{-1}}}

\def\la{\mathrel{\hbox{\rlap{\hbox{\lower4pt\hbox{$\sim$}}}{\raise2pt\hbox{$<$}}
}}}
\def\ga{\mathrel{\hbox{\rlap{\hbox{\lower4pt\hbox{$\sim$}}}{\raise2pt\hbox{$>$}}
}}}
\def\d25{$D_{25}$}

\begin{document}

\maketitle

\label{firstpage}

\begin{abstract}

We present a multi-mission X-ray analysis of a bright (peak observed 0.3--10 keV luminosity of $\sim 6 \times 10^{40}~{\rm erg~s^{-1}}$), but relatively highly absorbed ULX in the edge-on spiral galaxy NGC 5907.  The ULX is spectrally hard in X-rays ($\Gamma \sim 1.2$--$1.7$, when fitted with an absorbed power-law), and has a previously-reported hard spectral break consistent with it being in the ultraluminous accretion state.  It is also relatively highly absorbed for a ULX, with a column of $\sim 0.4 - 0.9 \times 10^{22} \rm ~atom~cm^{-2}$ in addition to the line-of-sight column in our Galaxy.  Although its X-ray spectra are well represented by accretion disc models, its variability characteristics argue against this interpretation.  The ULX spectra instead appear dominated by a cool, optically-thick Comptonising corona.  We discuss how the measured 9 per cent rms variability and a hardening of the spectrum as its flux diminishes might be reconciled with the effects of a very massive, radiatively-driven wind, and subtle changes in the corona respectively.  We speculate that the cool disc-like spectral component thought to be produced by the wind in other ULXs may be missing from the observed spectrum due to a combination of a low temperature ($\sim$ 0.1\,keV), and the high column to the ULX.  We find no evidence, other than its extreme X-ray luminosity, for the presence of an intermediate mass black hole ($\sim 10^2 - 10^4 \Msun$) in this object.  Rather, the observations can be consistently explained by a massive ($\ga 20 \Msun$) stellar remnant black hole in a super-Eddington accretion state.

\end{abstract}

\begin{keywords}
accretion, accretion discs -- black hole physics -- X rays: binaries -- 
X rays: galaxies
\end{keywords}

\section{Introduction}

Ultraluminous X-ray sources (ULXs) are X-ray sources with observed luminosities in excess of $10^{39}~{\rm erg~s^{-1}}$, which is equivalent to the Eddington luminosity for a $\sim 10 M_{\odot}$ black hole.  They are located outside the nuclear regions of galaxies, hence are very unlikely to host super-massive black holes (SMBHs, $M_{\rm BH} > 10^5 M_{\odot}$).  Instead, it seems increasingly likely that most ULXs are in fact X-ray binaries (XRBs), with stellar remnant black hole primaries \citep{feng_and_soria_2011}, although some intermediate-mass black hole candidates (IMBHs, $M_{\rm BH} \sim 10^2 - 10^4 M_{\odot}$) remain amongst the most luminous ULXs (e.g. \citealt{farrell_etal_2009}; \citealt{sutton_etal_2012}).  To produce the observed X-ray luminosities seen in many ULXs, stellar-mass black holes (sMBHs, $M_{\rm BH} < 20 M_{\odot}$; \citealt{feng_and_soria_2011}) must be accreting at apparently super-Eddington rates, rather than in any of the known sub-Eddington accretion states commonly seen in Galactic XRBs, and X-ray spectroscopy appears to confirm this (\citealt{roberts_2007}; \citealt*{gladstone_etal_2009}).  However, some of their luminosity could still be explained by the presence of massive stellar remnant black holes (MsBHs, $M_{\rm BH} \approx 20 - 100 M_{\odot}$; \citealt{feng_and_soria_2011}), that may be the end products of massive stellar evolution in low metallicity regions of the local Universe (\citealt{zampieri_and_roberts_2009}; \citealt{belczynski_etal_2010}).

The precise mechanisms of super-Eddington accretion are still far from well understood, although much theoretical progress has been made in recent years (see e.g. \citealt{dotan_and_shaviv_2011} and references therein).  Observationally, the high quality X-ray spectroscopy offered by the European Photon Imaging Camera (EPIC) detectors aboard {\it XMM-Newton\/} has revealed common morphologies for the best quality ULX spectra: a very broad, disc-like continuum for some ULXs, possibly close to the Eddington threshold, and a two-component spectrum consisting of a soft excess and a harder component with a distinct turnover at energies of several keV, which it has been suggested could be associated with higher super-Eddington accretion rates (e.g \citealt*{stobbart_etal_2006}; \citealt{goncalves_and_soria_2006}; \citealt{gladstone_etal_2009}).  Despite the high quality of the data, these spectra can be fitted with multiple physical models -- for instance, the disc-like spectra are well-fitted by advection-dominated `slim' disc models (e.g. \citealt*{watarai_etal_2001}; \citealt{isobe_etal_2012}), but may be preferentially fit by a combination of an advection-dominated disc and a cold, optically-thick corona in some cases (\citeauthor*{middleton_etal_2011b} 2011b; \citealt{middleton_etal_2012}).  The combination of a disc and a cool, thick corona may also provide a good explanation for the two component spectra seen in more luminous ULXs \citep{gladstone_etal_2009}, but the interpretation of these two components may not be completely straightforward -- for instance a massive, radiatively-driven wind offers a very plausible explanation for the soft component (\citealt{kajava_and_poutanen_2009}; \citeauthor{middleton_etal_2011a} 2011a).  Other interpretations are also possible, for example a combination of X-ray reflection and extremely broadened emission lines offers a statistically-valid solution for many ULXs (\citealt{caballero-garcia_and_fabian_2010}; \citeauthor{walton_etal_2011a} 2011a).

So despite the clear advances in understanding provided by X-ray spectroscopy of ULXs, it is clear that it does not currently offer unique solutions to the nature of ULXs.  More information is therefore required to break this degeneracy.  One means of addressing this is to examine the variation of the X-ray spectra with source luminosity, and so rule out physically implausible behaviour.  Spectral variations have long been reported for individual ULXs observed on multiple occasions, although the analyses of these data have generally been based on simple phenomenological spectral models due to the moderate quality of the data (e.g. \citealt*{kubota_etal_2002};  \citealt{dewangan_etal_2004}; \citealt{feng_and_kaaret_2006b}; \citealt{soria_etal_2009}).  A more systematic search for variability patterns was undertaken by both \cite{kajava_and_poutanen_2009} and \cite{feng_and_kaaret_2009}, on the basis of power-law continua and multi-colour disc blackbody models.  They both found a number of ULXs with correlated luminosity and power-law photon index, although \cite{feng_and_kaaret_2009} found others in starburst galaxies appeared considerably less variable in spectral shape at different fluxes.  \cite{kajava_and_poutanen_2009} also found several ULXs that were well-described by accretion disc spectra, and showed luminosity-temperature relationships consistent with this interpretation.  Perhaps most crucially, they found that some ULXs display spectral pivoting, consistent with two-component spectra; and the luminosity-temperature evolution of the soft component is consistent with an outflowing wind, not an accretion disc.

Further insights have come from studies of individual objects using high quality {\it XMM-Newton\/} data and more physically-motivated models.  However, these have posed more questions than they have answered.  NGC 1313 X-2 has been well-fitted with a model comprised of a Comptonising corona and an accretion disc, with the corona appearing cooler and optically thicker at higher luminosities \citep{pintore_and_zampieri_2012}.  In contrast,  \cite{kajava_etal_2012} report that when modelled by the same disc plus corona model, Ho II X-1 shows an opposite behaviour: an increase in coronal temperature with luminosity.  This is compounded by further, complex spectral variability in previous {\it Swift\/} monitoring campaigns \citep{grise_etal_2010}.  NGC 1313 X-1 is different again; it shows a range of coronal temperatures and optical depths that do not correlate with luminosity \citep{pintore_and_zampieri_2012}\footnote{ESO 243-49 HLX-1 is again different, showing spectral variability similar to Galactic sMBH binaries, albeit at higher X-ray luminosities and lower disc temperatures; and it has an inner disc temperature - X-ray luminosity relationship consistent with $L_{\rm X} \propto T^4$ (\citealt{servillat_etal_2011}; \citealt{godet_etal_2012}).  However, its unique behaviour marks ESO 243-49 HLX1 as physically distinct from the bulk of the ULX population, and so it is arguably a poor comparator for most ULXs in which there is no longer a strong suspicion of the presence of an IMBH.}.  From this small sample it therefore appears that the ULX population displays a variety of behaviours, possibly indicating some additional parameters -- such as accretion geometry and/or viewing angle and/or complex absorption -- are influential in their spectra.  Clearly further investigation is necessary to gain an understanding of the physics of the accretion states in ULXs.

In this work we present a thorough analysis of available observations of a ULX in the edge on spiral galaxy NGC 5907\footnote{NGC 5907 is an edge-on SA(s)c spiral galaxy with a prominent dust lane.  The regions west of the dust lane were originally catalogued as NGC 5906.}.  This object (2XMM J151558.6+561810, henceforth NGC 5907 ULX) was first identified as a ULX by \citeauthor{walton_etal_2011b} (2011b) in a cross correlation of the 2XMM-DR1 \citep{watson_etal_2009} and RC3 galaxy \citep{deVaucouleurs_etal_1991} catalogues. NGC 5907 ULX was sufficiently X-ray luminous in those observations (taken in 2003) to be included in the extreme ULX sample of \cite{sutton_etal_2012}; although unlike other ULXs in that sample, the identification of an X-ray  spectral break indicated that this source is likely in a super-Eddington regime.  However, in a 2010 {\it Swift\/} follow-up observation its flux was observed to have dropped substantially.  This source therefore provided an excellent opportunity to investigate spectral variations with flux in the ultraluminous state, and further {\it Swift\/} monitoring observations in 2011/2012 and {\it XMM-Newton\/} triggered observations in early 2012 (triggered by a factor of three diminution in count rate below the 2003 levels) were obtained.  In the remainder of this paper we constrain the changes in spectral and timing properties of NGC 5907 ULX with flux using the above data.  We then discuss these changes both in light of the various traits seen in other ULXs, and the physical models used to explain these objects.  We adopt a distance of $d = 13.4$ Mpc to NGC 5907 \citep{tully_etal_2009} throughout this work.

\section{Data reduction}

NGC 5907 ULX has been observed by a variety of X-ray observatories.  In addition to the {\it XMM-Newton\/} EPIC and {\it Swift\/} XRT observations introduced above, we also (somewhat fortuitously) obtained two proprietary {\it Chandra\/} ACIS-S observations within a couple of days of the latter of the two new {\it XMM-Newton\/} observations.  Further details of observations with all three observatories are provided in an X-ray observation log (Table \ref{obs_log}).  The source location was also observed during a series of {\it ROSAT} PSPC-B observations in 1992 (sequence ID RP600190N00), although the ULX was not detected. 

\begin{table}
\caption{X-ray observation log}
\centering
\begin{tabular}{cccc}
\hline
Date~$^a$ & Obs. ID~$^b$ & ${t_{\rm exp}}~^c$ & Off-axis angle~$^d$ \\
\hline

\multicolumn{4}{c}{{\it ROSAT} PSPC-B} \\
1992-01-05 & RP600190N00$^e$ & 18.1 & 1.14 \\

\multicolumn{4}{c}{{\it XMM-Newton} EPIC} \\
2003-02-20 & 0145190201 & 10.7/18.4/18.8 & 2.40 \\
2003-02-28 & 0145190101 & 10.6/17.2/17.5 & 2.41 \\
2012-02-05 & 0673920201 & 4.5/11.9/12.1 & 1.20 \\
2012-02-09 & 0673920301 & 12.2/17.3/16.3 & 1.12 \\

\multicolumn{4}{c}{{\it Swift} XRT} \\
2010-08-12 & 00031785001 & 5.6 & 2.42 \\
2010-08-13 & 00031785002 & 5.2 & 3.57 \\
2010-08-15 & 00031785003 & 6.8 & 1.78 \\
2011-05-04 & 00031785005 & 1.2 & 1.85 \\
2011-05-15 & 00031785006 & 1.9 & 1.55 \\
2011-05-29 & 00031785007 & 2.6 & 2.09 \\
2011-06-13 & 00031785008 & 2.1 & 2.20 \\
2011-11-20 & 00031785012 & 2.1 & 3.29 \\
2011-12-04 & 00031785013 & 0.9 & 2.74 \\
2011-12-18 & 00031785014 & 1.0 & 2.17 \\
2011-12-21 & 00031785015 & 0.9 & 1.31 \\
2012-01-01 & 00031785016 & 1.9 & 1.68 \\
2012-01-15 & 00031785017 & 1.7 & 0.65 \\
2012-01-29 & 00031785018 & 2.2 & 0.91 \\
2012-02-12 & 00031785019 & 1.5 & 3.61 \\

\multicolumn{4}{c}{{\it Chandra} ACIS-S} \\
2012-02-11 & 12987 & 16.0 & 0.02 \\
2012-02-11 & 14391 & 13.1 & 0.02 \\ 

\hline
\end{tabular}
\begin{minipage}{\linewidth}
Notes:
$^a$Observation start date (yyyy-mm-dd).
$^b$Observation identifier.
$^c$Cleaned {\it Chandra} ACIS-S/{\it Swift} XRT/{\it XMM-Newton} EPIC (pn/MOS1/MOS2)/combined {\it ROSAT} PSPC-B exposure time (ks).
$^d$Angle between the boresight position and the 2XMM co-ordinates of NGC 5907 ULX, in arcminutes.
$^e$ROSAT observation sequence ID, observations were taken over several days between 1992-01-05 -- 1992-01-12.
\end{minipage}
\label{obs_log}
\end{table}

The {\it XMM-Newton} data were reduced using {\it XMM-Newton} {\sc sas}\footnote{\tt http://xmm.esac.esa.int/sas/} version 11.0.0.  Both observations 0145190201 and 0145190101 were subject to severe background flaring.  This resulted in multiple exposures being obtained in all three EPIC detectors in 0145190201, and solely the MOS detectors in 0145190101.  However only one exposure per detector contained a significant duration of good (i.e low background) time, so this was identified and utilised in the following analysis.  To account for any background flares still present within these and the other datasets, we constructed good time interval (GTI) files from the high energy (10--15 keV) full field light curves, to exclude time intervals with count rates of typically $\ga 1 ~{\rm ct~s^{-1}}$ in the pn data, and $\ga 0.5 \rm ~ct~s^{-1}$ in the MOS.  Source energy spectra and light curves were extracted from circular apertures, with radii 30--40 arcseconds, centred on the position of NGC 5907 ULX.  Circular apertures of 40--50 arcsecond radius were used to extract background data, that were located a similar distance from the readout node on the neighbouring CCD to the one containing the source for the pn detector, and on the same CCD as the source for both of the MOS detectors.

Energy spectra and light curves were extracted from the data using the {\sc sas} task {\sc evselect}, which used both the source and background regions and the GTI files.  Standard event patterns ({\tt PATTERN} $\le 4$ for the pn, {\tt PATTERN} $\le 12$ for the MOS) were selected.  When extracting spectra the filter {\tt FLAG} $=0$ was imposed; {\tt \#xmmea\_ep} or {\tt \#xmmea\_em} were used instead for light curves.  Spectra were corrected using {\sc backscale} to calculate the areas of source and background extraction regions for subsequent normalisation; redistribution matrices and ancillary response files were extracted using {\sc rmfgen} and {\sc arfgen} respectively; finally spectral files were rebinned to a minimum of 20 counts per bin using {\sc specgroup}.

Pipeline-reduced {\it Swift} data for observations of NGC 5907 ULX  were obtained from the NASA HEASARC archive\footnote{\tt http://heasarc.gsfc.nasa.gov/docs/archive.html}.  Source events were extracted from circular apertures centred on the X-ray source, with radii of 20 pixels ($\sim$ 47 arcseconds); for background events annular apertures were used, again centred on the ULX, with inner and outer radii of 50 \& 120 pixels around the source positions.  Source and background spectra were extracted using {\sc xselect}\footnote{\tt http://heasarc.gsfc.nasa.gov/docs/software/lheasoft/\\ftools/xselect/} v2.4.  The response file {\tt swxpc0to12s6\_20070901v011.rmf} was used in spectral fitting, and ancillary response files were extracted using {\sc xrtmkarf}.  {\it Swift} spectra were not grouped, instead Cash statistics \citep{cash_1979} were used for subsequent spectral fitting, as described below.

{\it Chandra} data were reduced using {\sc ciao}\footnote{\tt http://cxc.harvard.edu/ciao/} version 4.4 and calibration files from the \chan {\sc caldb} version 4.4.8.  Both of the {\it Chandra} observations were taken using the 1/4 subarray to mitigate against the effects of pile-up, and the pile-up fraction in the actual observations was estimated as $\sim 2$ per cent using {\sc pimms}.  Background lightcurves were extracted from the ACIS-S3 chip, with all of the detected point sources removed; these were then checked for any flaring using the {\sc lc\_sigma\_clip} routine\footnote{\tt http://cxc.harvard.edu/ciao/ahelp/lc\_sigma\_clip.html}.  No further filtering of background flares was necessary for either observation.  Circular apertures with radii of 5 arcseconds (corresponding to an on-axis fractional encircled energy of $\ga 99$ per cent at 1 keV) were used to extract the source data.  Annular apertures, centred on the source, were used to characterise the background count rate for each observation.  These had inner and outer radii of 13--17 and 16--18 pixels for observations 12987 and 14391 respectively, set to exclude faint point sources visible in the sensitive ACIS-S images.  The data products were then extracted from the level 2 event files, to include all standard good event detections with energies between 0.3--10 keV, within the source data extraction apertures.  The \chan spectra were extracted, and grouped with $\ga 20$ counts per energy bin using the {\sc ciao} script {\sc specextract}, which also produces the appropriate response matrices for spectral analysis.

\section{Data analysis \& results}

\subsection{Spatial analysis}

In \cite{sutton_etal_2012} we identified a bright, blue optical source within the {\it XMM-Newton} positional uncertainty region for NGC 5907 ULX in Hubble legacy archive{\footnote {\tt http://hla.stsci.edu/}} (HLA) images.  Here we repeat this analysis using the superior astrometry and smaller uncertainty region provided by the recently-obtained {\it Chandra} data.  We firstly attempted to improve the astrometry in the {\it Chandra} image, by matching other bright X-ray point sources detected in {\it Chandra} observation 12987 with an optical catalogue. The {\sc ciao} script {\sc wavdetect} was used to produce a list of X-ray sources detected within 6 arcminutes of NGC 5907 ULX, and we checked for optical counterparts in SDSS-II DR7 to sources with greater than 15 net X-ray counts.  The SDSS catalogue was chosen primarily because it had been used by the HLA pipeline to correct the astrometry of the {\it HST} data, and when checked was found to provide an acceptable solution.  However, only a single {\it Chandra} source had a clear counterpart in the SDSS; we therefore carried out an astrometric correction using this object, but stress that it is particularly tentative.  We also extracted the {\it Chandra} position of NGC 5907 ULX using {\sc wavdetect}, which combined with the raw astrometric precision of {\it Chandra} is 15:15:58.63(0.02)+56:18:10.4(0.4) (the RA and dec errors shown in brackets are the $1 \sigma$ uncertainties, based on a combination of the centroiding error and the error on the astrometric solution for the {\it Chandra} field); the `tentatively-corrected' {\it Chandra} position of the ULX is 15:15:58.62(0.01)+56:18:10.3(0.1).  We then combined the new astrometric uncertainty of the X-ray data and the X-ray source centroid with the {\it HST} astrometric precision (as per \citealt*{roberts_etal_2008}) to calculate a new 90 per cent uncertainty region for the source position.  Even at the high astrometric precision of {\it Chandra}, NGC 5907 ULX at least marginally retains the bright blue optical counterpart in the HLA images \citep{sutton_etal_2012} - it lies within the uncorrected 90 per cent uncertainty region, and on the outer edge of the 90 per cent uncertainty region after the tentative correction. 

The counterpart has previously been reported with $m_{\rm F450W} = 21.5$ (\citealt{sutton_etal_2012}; corrected for Galactic extinction). Here we report an estimated magnitude limit in the F814W band, which was calculated by subtracting an estimate of the background flux in an annulus centred on the target position from the upper limit of flux at the location of the optical source. We then converted this to a magnitude and corrected for Galactic extinction, obtaining a $3 \sigma$ limit of $m_{\rm F814W}>24$. If the {\it HST} magnitudes are used as an approximation of Johnson magnitudes, this implies an unphysical colour of $B-I \la -3.5$.  This therefore questions the veracity of the HLA detection, so as a check we obtained the individual flat-field calibrated distorted images corresponding to each exposure from the Barbara A. Mikulski Archive for Space Telescopes{\footnote {\tt http://archive.stsci.edu/}} (MAST), and aligned the images in each filter using {\sc tweakreg} in the {\it HST} {\sc drizzlepac} software package{\footnote {\tt http://www.stsci.edu/hst/HST\_overview/drizzlepac}}.  When we combined the individual images using {\sc astrodrizzle} we were able to reproduce the detection of a bright source in the F450W band, as seen in the HLA images.  However, upon examining the 3 individual F450W exposures we found detections close to the position of the counterpart in only 2 of the 3 images.  Given the small temporal separation of the 3 exposures (which were all taken between 1996-03-01 21:44:16 and 22:08:36), it seems unlikely that a variable counterpart could be the cause of the non-detection.  Instead, we suggest that the counterpart previously identified in the HLA images was a spurious detection, probably caused by the presence of overlapping cosmic rays in 2 of the 3 exposures.  This suggests a potential issue with the cosmic ray rejection during the HLA processing.

\subsection{X-ray spectral analysis}

\begin{table}
\caption{X-ray spectral analysis: power-law}
\centering
\begin{tabular}{ccccc}
\hline
Obs. ID & $\chi^2 / \rm{dof}~^a$ & ${N_{\rm H}}~^b$ & $\Gamma~^c$ & Norm.$~^d$  \\
\hline
0145190201 & 268.4/335 & $0.85 \pm 0.06$ & $1.65 \pm 0.06$ & $3.5 \pm 0.3$ \\
0145190101 & 272.5/256 & $0.82 \pm 0.07$ & $1.65\pm 0.07$ & $2.8 \pm 0.3$ \\
0673920201 & 63.5/77 & $0.5 \pm 0.1$ & $1.2 \pm 0.1$ & $0.8^{+0.2}_{-0.1}$ \\
0673920301 & 169.5/139 & $0.7 \pm 0.1$ & $1.34^{+0.09}_{-0.08}$ & $0.9^{+0.2}_{-0.1}$ \\
12987 & 29.9/32 & $0.8 \pm 0.3$ & $1.2 \pm 0.2$ & $0.8^{+0.3}_{-0.2}$ \\
14391 & 34.1/26 & $0.9^{+0.4}_{-0.3}$ & $1.2 \pm 0.3$ & $0.9^{+0.4}_{-0.3}$ \\
\hline
\end{tabular}
\begin{minipage}{\linewidth}
Notes:
$^a$Statistical goodness of the fit, in-terms of the $\chi^2$ statistic and number of degrees of freedom.
$^b$Absorption column density external to our Galaxy ($\times 10^{22}~\rm{cm}^{-2}$).
$^c$Power-law photon index.
$^d$Normalisation of the power-law spectral model ($\rm \times 10^{-4}~photons~keV^{-1}~cm^{-2}~s^{-1}$).
\end{minipage}
\label{pow}
\end{table}

There were sufficient counts available to enable X-ray spectral analysis in all of the {\it XMM-Newton} and {\it Chandra} observations, so these were fitted with spectral models using {\sc xspec} v.12.6.0.  Firstly, the spectra were modelled using a simple, unphysical single component model -- a doubly absorbed power-law ({\tt tbabs $\times$ tbabs $\times$ powerlaw} in {\sc xspec}), with the abundances set to the values of \cite*{wilms_etal_2000}.  The first absorption component was frozen to the Galactic column in the direction of NGC 5907 ULX ($1.38 \times 10^{20}~{\rm cm^{-2}}$, \citealt{dickey_and_lockman_1990}), and the second was left free to model absorption intrinsic to the source and/or its host galaxy.  An additional multiplicative constant was included when fitting the {\it XMM-Newton} spectra, which was frozen to unity for the pn channel, and free to vary for the two MOS channels to account for uncertainties in the calibration between the EPIC detectors.  The parameters resulting from this spectral fitting are shown in Table~\ref{pow}.  Given that a power-law is an empirical representation of the data, rather than a physical model, its uses are limited.  Indeed, its use for this data is inherently questionable given the previous demonstration by \cite{sutton_etal_2012}\footnote{We note that further examination of this previous analysis indicated that the background emission was over-subtracted from the spectrum of NGC 5907 ULX in that work.  However, we repeated the analysis of \cite{sutton_etal_2012} for the newly processed data, and found that the break was still detected in the 2--10 keV spectrum extracted from both of the high flux {\it XMM-Newton} observations, at greater than $3 \sigma$ significance.} of the subtle but statistically significant presence of a high energy turnover\footnote{Although it appears contradictory that a power-law model can provide a reasonable fit to the full band data when a break is present at high energies, this is also the case in 3/12 objects looked at by \citet{gladstone_etal_2009}, where the data quality was similar to that presented here.  In each of those cases inspection of the data above 2 keV showed a statistically robust detection of a break.}, indicative of ultraluminous state spectra, in the data from the 2003 {\it XMM-Newton\/} observations.  However, a power-law still provides a reasonable approximation to the 0.3--10 keV X-ray spectra in all cases (cf. Table 2), and proves useful in two ways.  Firstly, as the data from the individual {\it Swift} monitoring observations had insufficient statistics for spectral fitting, we used these fits to provide an estimate of the source flux in those epochs (see below).  Secondly, it provides a strong indication of spectral variability in the ULX.  Although the absorption column remains roughly constant throughout the six spectral datasets, the slope shows a marked change between the 2003 and 2012 data, hardening as the normalisation (and so the flux) decreases.

\begin{table*}
\caption{X-ray spectral analysis: physical models}
\centering
\begin{tabular}{lccccccc}
\hline
%------------------------------------------------------------------------------------------------
\multicolumn{8}{c}{{\tt tbabs $\times$ tbabs $\times$ diskbb}} \\
Group ID & $\chi^2 / \rm{dof}~^a$ & ${N_{\rm H}}~^b$ & ${kT_{\rm in}}~^c$ & Norm~$^d$ & & & \\
%\hline
High flux & 550.8/593 & $0.48 \pm 0.03$ & $2.03 \pm 0.08$ & 
$5.7^{+0.8}_{-0.7} \times 10^{-3}$ & & & \\
Low flux & 311.0/280 & $0.46^{+0.05}_{-0.06}$ & $2.9^{+0.3}_{-0.2}$ & $8 
\pm 2 \times 10^{-4}$ & & & \\
\\
%------------------------------------------------------------------------------------------------
\multicolumn{8}{c}{{\tt tbabs $\times$ tbabs $\times$ diskpbb}} \\
Group ID & $\chi^2 / \rm{dof}~^a$ & ${N_{\rm H}}~^b$ & ${kT_{\rm in}}~^c$ & $p~^e$ & Norm~$^d$ & & \\
High flux & 511.9/592 & $0.68^{+0.07}_{-0.05}$ & $2.9^{+0.8}_{-0.3}$ & 
$0.61 \pm 0.02$ & $<0.001$ & & \\
Low flux & 308.7/279 & $0.6 \pm 0.1$ & $3.7^{+3.6}_{-0.9}$ & 
$0.68^{+0.07}_{-0.05}$ & $<7 \times 10^{-4}$ & & \\
\\
%------------------------------------------------------------------------------------------------
\multicolumn{8}{c}{{\tt tbabs $\times$ tbabs $\times$ comptt}} \\
Group ID & $\chi^2 / \rm{dof}~^a$ & ${N_{\rm H}}~^b$ & ${kT_{0}}~^f$ & $kT~^g$ & $\tau~^h$ & Norm~$^d$ & \\
High flux & 503.5/591 & $0.47^{+0.12}_{-0.09}$ & $0.31^{+0.05}_{-0.06}$ 
& $1.9 \pm 0.2$ & $9.8 \pm 0.9$ & $3.0^{+0.5}_{-0.4} \times 10^{-4}$ & \\
Low flux & 306.0/278 & $0.2^{+0.2}_{-0.1}$ & $0.45^{+0.09}_{-0.12}$ & 
$2.6^{+3.1}_{-0.6}$ & $9^{+2}_{-4}$ & $<1 \times 10^{-4}$ & \\
\\
%------------------------------------------------------------------------------------------------
\multicolumn{8}{c}{{\tt tbabs $\times$ tbabs $\times$ (diskbb$+$comptt)}~$^i$} \\
Group ID & $\chi^2 / \rm{dof}~^a$ & ${N_{\rm H}}~^b$ & ${kT_{\rm in}}~^{c,j}$ & Norm~$^d$ & $kT~^g$ & $\tau~^h$ & Norm~$^d$ \\
High flux & 503.2/591 & $0.55^{+0.17}_{-0.09}$ & 0.3 & $<2$ & $1.9 \pm 
0.2$ & $9.9 \pm 0.8$ & $3.1^{+0.2}_{-0.3} \times 10^{-4}$ \\
Low flux & 310.4/278 & $0.44^{+0.20}_{-0.09}$ & 0.3 & $<0.2$ & 
$2.1^{+0.5}_{-0.3}$ & $11 \pm 2$ & $1.1 \pm 0.1 \times 10^{-4}$ \\
High flux & 504.9/591 & $0.84^{+0.09}_{-0.08}$ & 0.1 & 
$1000^{+1400}_{-900}$ & $2.1^{+0.3}_{-0.2}$ & $8.6 \pm 0.9$ & $6.3 \pm 0.6 \times 10^{-4}$ \\
Low flux  & 306.2/278 & $0.8^{+0.2}_{-0.1}$ & 0.1 & $800^{+1500}_{-700}$ 
& $2.5^{+1.5}_{-0.5}$ & $9^{+2}_{-3}$ & $1.8^{+0.3}_{-0.5} \times 
10^{-4}$ \\
\hline
%------------------------------------------------------------------------------------------------
\end{tabular}
\begin{minipage}{\linewidth}
Notes: Results from fitting various spectral models to the flux-binned {\it XMM-Newton} and {\it Chandra} spectra: the bright flux bin includes observations 0145190201 and 0145190101; the faint flux bin includes observations 0673920201, 0673920301, 12987 and 14391.  The errors and limits shown are the 90 per cent confidence ranges.
$^a$Statistical goodness of the fit, in terms of the $\chi^2$ statistic and number of degrees of freedom.
$^b$Absorption column density external to our Galaxy ($\times 10^{22}~\rm{cm}^{-2}$).
$^c$Inner-disc temperature (keV).
$^d$Normalisation of the spectral component.  For the disc models this is $(R_{\rm in}/D)^2 \rm ~cos(\theta)$ for a disc with inner radius $R_{\rm in}$ (in km) at a distance $D$ (in units of 10 kpc) and at an inclination $\theta$.
$^e$Exponent of the radial dependence of temperature in the {\it p}-free disc model.
$^f$Input soft photon temperature (keV) of the Comptonising corona.
$^g$Plasma temperature (keV) of the Comptonised component.
$^h$Optical depth of the Comptonised component.
$^i$In the multi-colour disc plus Comptonising corona model the input soft photon temperature of the corona was set equal to the inner disc temperature.
$^j$The inner disc temperatures were fixed in value, as per the text.
\end{minipage}
\label{spec_par}
\end{table*}

\begin{figure}
\centering
\includegraphics[width=8.5cm]{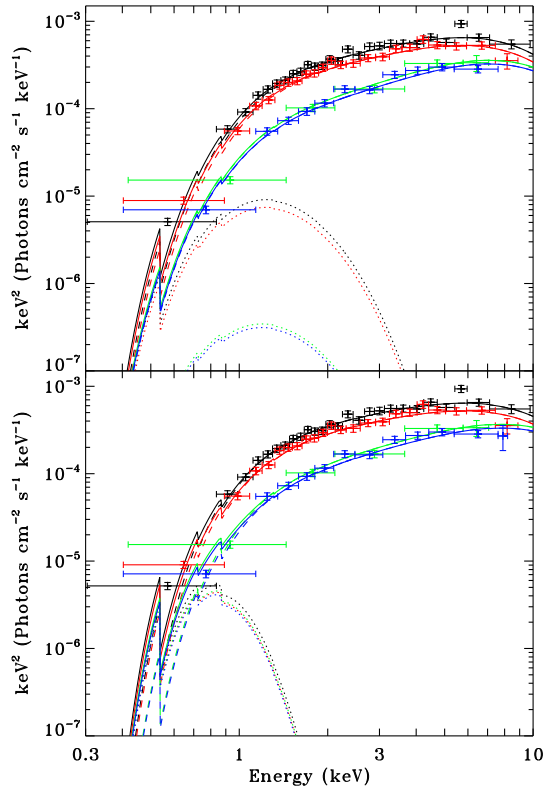}
\caption{The best fitting absorbed multi-colour disc (dotted lines) plus Comptonisation (dashed lines) models.  These models ({\tt tbabs $\times$ tbabs $\times$ (diskbb$+$comptt}) in {\sc xspec}) are fitted to the bright (0145190201 in black and 0145190101 in red) and faint (0673920201 in green and 0673920301 in blue; although used in the fitting, the {\it Chandra} spectra are not shown for clarity) flux binned {\it XMM-Newton} energy spectra.  Only EPIC pn data are shown, and data have been rebinned to $10 \sigma$ significance per data point.  The top panel shows the fit with the inner disc temperature and coronal seed temperature fixed to 0.3 keV; the bottom panel shows it fixed to 0.1 keV. The other model parameters of these fits are shown in Table~\ref {spec_par}.}
\label{spec_plot}
\end{figure}

So, in order to investigate the reasons for this spectral hardening, we attempted to fit some more physically motivated spectral models to the {\it XMM-Newton} and {\it Chandra} data.  To get the best possible constraints on the model parameters, and motivated by the apparent consistency in power-law spectra within the two separate periods of observations, we separated the spectra into a high flux group from the February 2003 observations ({\it XMM-Newton} observations 0145190201 and 0145190101), and a low flux group from February 2012 ({\it XMM-Newton} observations 0673920201 and 0673920301, plus {\it Chandra} observations 12987 and 14391).  Each group of spectra was fitted simultaneously, with all model parameters held fixed between the observations.  Similarly to the power-law, in each spectral model we included two absorption components ({\tt tbabs} in {\sc xspec}), one fixed to the Galactic column density in the direction of NGC 5907, the other free to model intrinsic absorption in the source/host galaxy.  We also included a multiplicative constant to allow for calibration uncertainties between the detectors and differences between observations; this was fixed to 1 for the {\it XMM-Newton} EPIC pn detector in one observation per flux group (0145190201 and 0673920201), and was left free to vary for other observations/detectors.  For the high flux bin, the constants for observation 0145190101 were $\sim$ 20 per cent lower than for 0145190201, but the constants of detectors within individual observations were all within 10 per cent.  For the faint flux bin, the constants of the {\it XMM-Newton} EPIC and {\it Chandra} ACIS-S detectors in each observation were within 10 per cent of each other.

Given the shape of the high flux spectra shown in Figure~4 of \cite{sutton_etal_2012} -- smooth, absorbed continua, with a turnover above 5 keV -- we began by fitting accretion disc models to the data.  A multi-colour disc blackbody ({\tt diskbb} in {\sc xspec}) model was initially employed, and resulted in statistically acceptable fits, but with rather unrealistically high inner disc temperatures ($kT_{\rm in} > 2$ keV, Table~\ref{spec_par}, first model).  We also attempted to fit the spectral data with a $p$-free disc model (Table~\ref{spec_par}, second model), which allows different values for the exponent of the radial dependence of temperature in the accretion disc {\it p}, with $p=0.75$ indicating a standard Shakura-Sunyaev disc, and $p \approx 0.5$ indicating the presence of a slim (advection-dominated) disc (cf. \citealt{vierdayanti_etal_2006}).  This resulted in $\Delta \chi^2 =$ 38.9 and 2.3 compared to the multi-colour disc fits, for one additional degree of freedom for the high and low flux grouped observations respectively, i.e. indicating a significantly improved fit for the high flux group.  However, in both cases the $p$-value was intermediate between the slim and standard disc solutions (with a standard disc acceptable within errors for the low flux group), and the disc remained anomalously hot, with $kT_{\rm in} \sim 3 - 4$ keV.  Furthermore, the normalisations of the $p$-free disc model were relatively unconstrained, with only upper limits present at the 90\% level.

We also attempted to fit the X-ray spectra with a Comptonising corona ({\tt comptt} in {\sc xspec}), as an optically thick, cool corona could provide the high energy spectral break \citep{sutton_etal_2012}.  Such a corona provided acceptable fits to both of the observation groups (Table \ref{spec_par}, third model), with slightly improved statistics compared to the $p$-free disc model ($\Delta\chi^2 = 2 - 8$ for one additional degree of freedom).  The physical parameters derived from the model are consistent with those observed in many other ULXs (cf. \citealt{gladstone_etal_2009}), implying a cool, optically-thick corona ($kT \sim 2 - 3$ keV; $\tau \sim 9 - 10$), with a cool medium providing the seed photons ($kT_0 \sim 0.31 - 0.45$).

Clearly, a spectrum consisting of a Comptonising corona with no accompanying emission from the input photon source is physically challenging (although scenarios in which a corona completely envelopes the X-ray emitting disc are certainly not impossible).  We therefore attempted to fit the spectra with a two component model consisting of a multi-colour disc plus a Comptonising corona ({\tt diskbb$+$comptt} in {\sc xspec}), with the disc assumed to provide the seed photons for the corona and hence the input soft photon temperature of the corona fixed to the inner disc temperature\footnote{This is a common approximation for the moderate quality spectra obtained from ULXs, although see \cite{pintore_and_zampieri_2012} for a discussion on the limitations of this approach.}.  However, this resulted in very poor constraints on the model parameters, and no improvement to the fit statistics.

Despite this model not being a statistical requirement of fits to the X-ray data of NGC 5907 ULX, cool disc-like spectral components are a very common feature in other similarly luminous ULXs with high quality spectral data, so we consider the constraints on whether such a feature is present here.  As a start point we note that the high absorption column to this object might act to mask the emission of an intrinsically soft spectral component; hence we attempted to estimate the amount of disc-like emission that could be intrinsically present, but obscured by the high column density, by fixing the disc temperature to a characteristic value.  While a free fit does not constrain this temperature, and fixing the temperatures reveals similar $\chi^2$ values for a range of disc temperatures between 0.1 and 1 keV, we choose to use 0.1 and 0.3 keV.  This is motivated by the range of cool disc temperatures seen in other, similar ULX fits e.g. \citeauthor{gladstone_etal_2009} (2009, their Table~8; they model the disc using the {\tt diskpn} model in {\sc xspec}, but state that this differs from {\tt diskbb} by less than 5 per cent for the same disc temperature); the low input photon temperature required by the Comptonisation-only model; and the hypothesis that a softer disc might more easily be hidden behind the high absorption column.   The results for these fits are shown in Table \ref{spec_par} (fourth model).  Clearly, the physical parameters of the coronal component are consistent within errors for both assumed cool disc temperatures; the main difference between the models for each disc temperature is that the cooler (0.1 keV) disc requires a substantially higher absorption column ($\sim 8 \times 10^{21} \rm ~cm^{-2}$, nearly twice that required for the 0.3 keV disc), which has the added implication that the model normalisations are all substantially larger for the cooler disc case.

We show the best-fitting disc plus corona models for the {\it XMM-Newton\/} data, unfolded from the instrument response, for both fixed disc temperatures in Figure~\ref{spec_plot}.  The plots clearly show why the soft disc is not required statistically; in both cases it contributes minimally to the observed spectrum.  Indeed, in 3 of 4 cases the disc components only produce $\la 0.5$ per cent of the observed 0.3-10\,keV flux, and the other case (high flux, 0.3 keV disc) has only $\sim 2.5$ per cent of its observed flux in the disc component.  However, the fraction of intrinsic (i.e. deabsorbed) flux in the disc-like component could be substantially higher for both disc temperatures.  We illustrate this in Table~\ref{cflux}, where we provide the deabsorbed (i.e. intrinsic) luminosities for the two components for both flux groups, and both assumed disc temperatures.  The deabsorbed  luminosities were calculated using the {\tt cflux} convolution model in {\sc xspec}, over the 0.3--10 keV energy range, with the appropriate normalisation frozen as required by the model.  Although the errors on the 0.1\,keV disc are very substantial due to the large extrapolation necessary to correct for the high absorption column, it is apparent that such a disc might have a very substantial contribution to the 0.3-10\,keV flux (at least $\sim 10$ per cent, and plausibly much higher), that is hidden from our view due to the high column.  The contribution is less for the 0.3\,keV disc, but even then up to a few per cent of the intrinsic flux may be in such a component.

Interestingly, these fits also provide a suggestion for the origin of the spectral changes.  In the case of the 0.3\,keV disc, it appears that the hardening of the spectrum is a result of the flux in the disc diminishing at a faster rate than the flux from the corona as the overall flux drops.  However, in the case of a 0.1\,keV disc this component appears to be consistent between the two observations, and the change in spectrum is originating in a slight (although not statistically distinctive given the quality of data) heating of the corona as the overall flux drops.

\begin{table}
\caption{Intrinsic luminosities for the components of the multi-colour disc plus corona model}
\centering
\begin{tabular}{lccc}
\hline
Group ID & $kT_{\rm in}$ & ${L_{\rm X,diskbb}}~^a$ & ${L_{\rm X,comptt}}~^b$ \\
\hline
High flux & 0.3 & $<0.5$ & $5.1 \pm 0.1$ \\
Low flux  & 0.3 & $<0.06$ & $2.3 \pm 0.1$ \\
High flux & 0.1 & $1.4^{+1.1}_{-0.8}$ & $6.3 \pm 0.2$ \\
Low flux  & 0.1 & $1.2^{+1.1}_{-0.7}$ & $2.7 \pm 0.2$ \\\hline
\end{tabular}
\begin{minipage}{\linewidth}
Notes: Errors and limits shown on fluxes are $1 \sigma$ confidence ranges.
$^a$0.3--10 keV intrinsic (i.e. deabsorbed) luminosity ($\times 10^{40} {\rm ~erg~s^{-1}}$) of the 0.3 keV multi-colour disc component of the spectral model.
$^b$Same as $^a$, but for the Comptonising corona.
\end{minipage}
\label{cflux}
\end{table}

\subsection{Long-term X-ray timing} 

\begin{figure*}
\centering
\includegraphics[width=12cm]{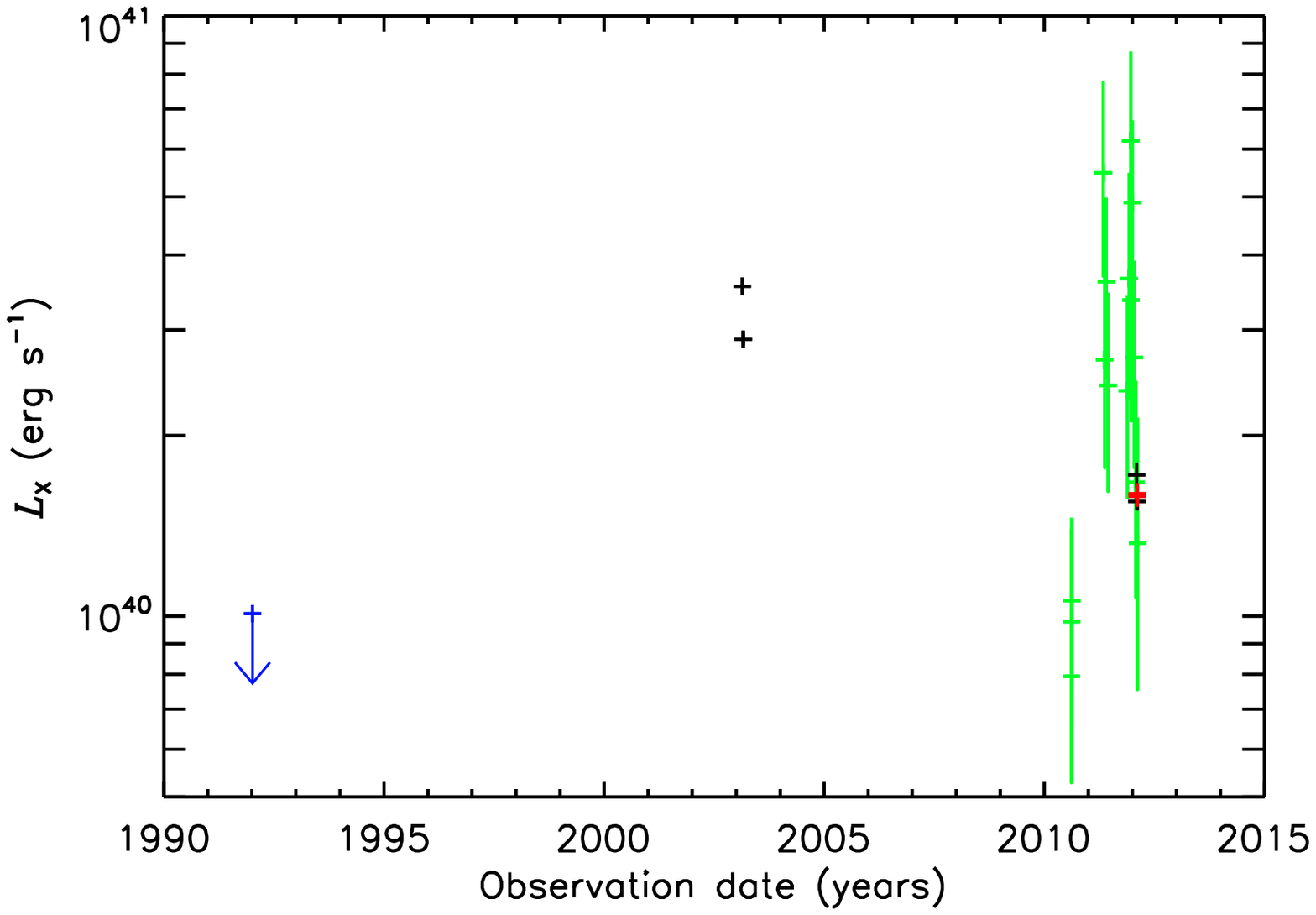}
\includegraphics[width=12cm]{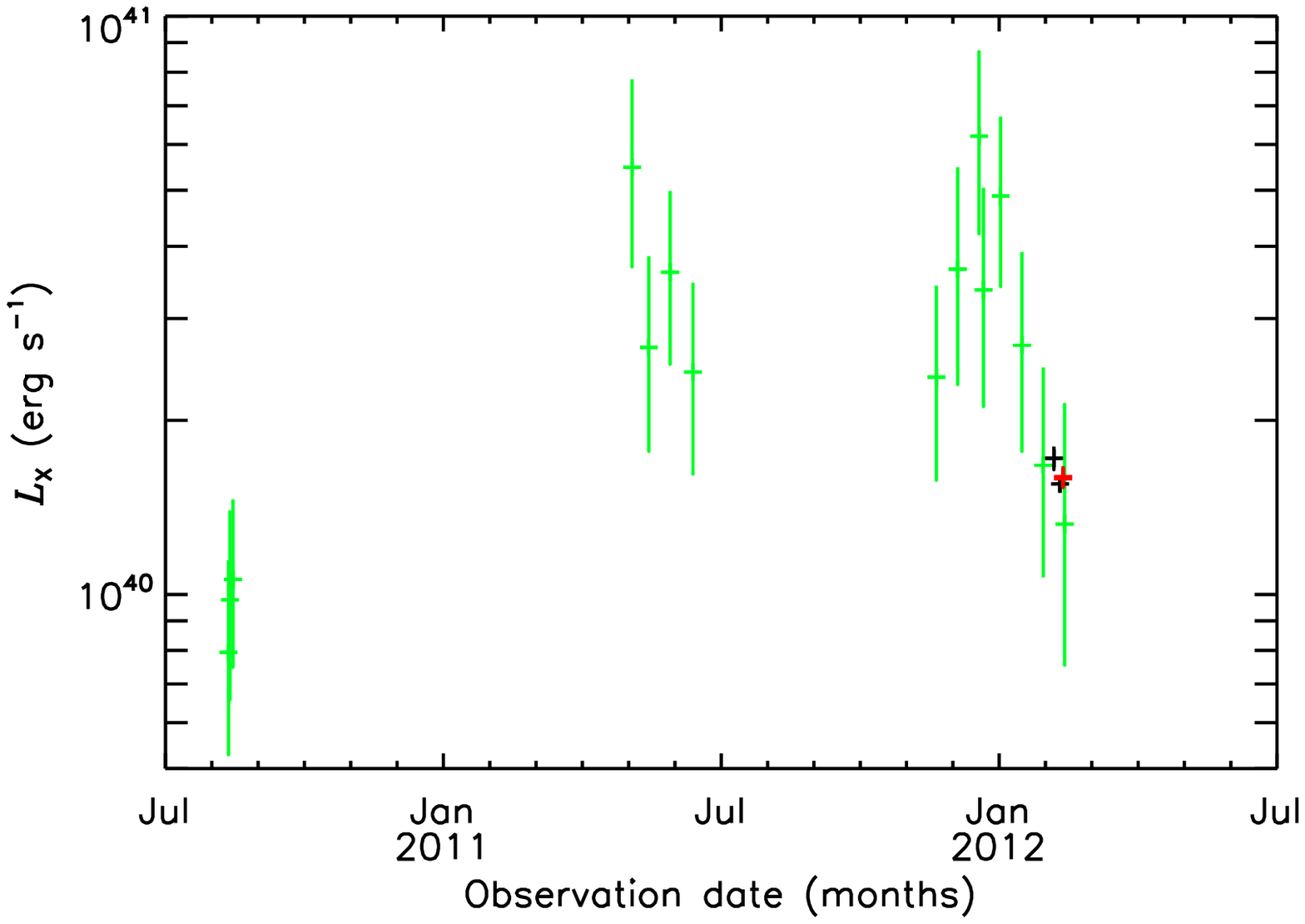}
\caption{Multi-mission long-term X-ray light curves of NGC 5907 ULX showing the observed 0.3--10\,keV fluxes and their associated $1 \sigma$ error bars.  Both the full range ({\it top}), and an enlargement displaying data from 2010 to present ({\it bottom}), are shown.  {\it XMM-Newton} fluxes are shown in black and {\it Chandra} in red; these were calculated using the best fitting absorbed Comptonisation spectral model for each group, with the normalisations for the individual observations free to vary.  {\it Swift} fluxes are shown in green; these are a mean of the two fluxes calculated using both the $\Gamma=1.7$ and $\Gamma=1.0$ spectral models, and the errors show the full $1 \sigma$ uncertainty range for both models (see Table \ref{flux}).  A $1 \sigma$ upper limit from an earlier series of {\it ROSAT} observations is shown in blue.}
\label{lightcurves}
\end{figure*}

\begin{table}
\caption{Observed 0.3--10 keV X-ray fluxes}
\centering
\begin{tabular}{ccc}
\hline
Obs. ID & \multicolumn{2}{c}{${f_{\rm X}}~^a$} \\
\hline

\multicolumn{3}{c}{{\it XMM-Newton}~$^b$} \\
0145190201 & \multicolumn{2}{c}{$16.5 \pm 0.3$} \\
0145190101 & \multicolumn{2}{c}{$13.5 \pm 0.3$} \\
0673920201 & \multicolumn{2}{c}{$8.0 \pm 0.4$} \\
0673920301 & \multicolumn{2}{c}{$7.2 \pm 0.2$} \\

\multicolumn{3}{c}{{\it Swift}~$^c$} \\
 & $\Gamma = 1.7$ & $\Gamma = 1.0$ \\
00031785001 & $3.0^{+0.7}_{-0.6}$ & $4.4^{+1.0}_{-0.9}$ \\
00031785002 & $3.7^{+0.8}_{-0.7}$ & $5 \pm 1$ \\
00031785003 & $4.1 \pm 0.7$ & $5.8^{+1.0}_{-0.9}$ \\
00031785005 & $21 \pm 4$ & $30^{+6}_{-5}$ \\
00031785006 & $10 \pm 2$ & $15 \pm 3$ \\
00031785007 & $14 \pm 2$ & $20 \pm 3$ \\
00031785008 & $9 \pm 2$ & $14^{+3}_{-2}$ \\
00031785012 & $9 \pm 2$ & $13^{+3}_{-2}$ \\
00031785013 & $14^{+4}_{-3}$ & $20^{+5}_{-4}$ \\
00031785014 & $23 \pm 4$ & $35 \pm 6$ \\
00031785015 & $13 \pm 3$ & $19^{+5}_{-4}$ \\
00031785016 & $18^{+3}_{-2}$ & $27 \pm 4$ \\
00031785017 & $10 \pm 2$ & $15 \pm 3$ \\
00031785018 & $6 \pm 1$ & $9 \pm 2$ \\
00031785019 & $5^{+2}_{-1}$ & $7^{+3}_{-2}$ \\

\multicolumn{3}{c}{{\it Chandra}~$^d$} \\
12987 & \multicolumn{2}{c}{$7.4 \pm 0.3$} \\
14391 & \multicolumn{2}{c}{$7.4 \pm 0.3$} \\

\hline
\end{tabular}
\begin{minipage}{\linewidth}
Notes: 
$^a$Observed 0.3--10 keV flux ($\times 10^{-13}~{\rm erg~cm^{-2}~s^{-1}}$), calculated for each mission as described below.  Errors shown for fluxes are the $1 \sigma$ confidence ranges.
$^b${\it XMM-Newton} EPIC fluxes are calculated using the absorbed multi-colour disc plus Comptonisation spectral model, and the {\tt cflux} convolution model in {\sc xspec}.
$^c${\it Swift} XRT fluxes were calculated for two different absorbed power-law spectral models, with an extra-Galactic absorption column of $N_{\rm H} = 8 \times 10^{21}~{\rm cm^{-2}}$ and photon indexes of $\Gamma = 1.7$ (the upper limit of the range seen in the brighter/softer {\it XMM-Newton} observations) and $\Gamma = 1.0$ (the lower limit of the range seen in the fainter/harder {\it XMM-Newton} observations).  All model components were frozen, except for the normalisation.
$^d${\it Chandra} ACIS-S fluxes were calculated as per the {\it XMM-Newton} data.
\end{minipage}
\label{flux}
\end{table}

In order to constrain the long-term variability of NGC 5907 ULX, we extracted a flux from each observation of the source.  However, the exact method used differed for data from each observatory.  There were insufficient counts per observation in the {\it Swift} spectra for statistically valid spectral model fitting, so parameters from the absorbed power-law models fitted to the {\it XMM-Newton} data were used instead.  A typical value of intrinsic absorption of $8 \times 10^{21}~{\rm cm^{-2}}$ was assumed.  Even though the power-law fit parameters in Table~\ref{pow} indicated that the source became spectrally harder at fainter fluxes in the small number of high quality X-ray observations, such behaviour was not assumed to be universally present in the source.  Instead two values of $\Gamma$ were used, $\Gamma = 1.0$ and 1.7, that approximate the upper and lower limits of the range observed by {\it XMM-Newton\/} and {\it Chandra\/}.  Fluxes were obtained by fitting both absorbed power-law continua to the {\it Swift} data in {\sc xspec}, using the C-statistic, with only the normalisations free to vary.  The resulting fluxes are shown in Table \ref{flux}.  As higher quality spectral data were available for the {\it XMM-Newton} and {\it Chandra} observations, fluxes and errors were obtained from the absorbed Comptonisation model, using the best fit parameters for each flux group but with the normalisation of each observation allowed to vary.  These values are also shown in Table~\ref{flux}, and were calculated using the {\sc xspec} {\tt cflux} convolution model component.

We also searched for any earlier detections of NGC 5907 ULX in archival {\it ROSAT} data.  The source location fell within the field of view of a series of {\it ROSAT} PSPC-B exposures, taken between 1992-01-05 and 1992-01-12, with the source position at an off-axis angle of 1.14 arcminutes, and with a total exposure time of 18.1 ks.  However, there was no significant detection of an X-ray source at the location of the ULX.  Instead, we estimate an upper limit by counting the number of photons in a 1 arcminute circular region centred on the 2XMM position of the ULX, and we estimate the background using 3 such circles placed close to the source region.  The upper limit on count rate was converted to a flux limit and extrapolated to the 0.3--10 keV band using {\sc pimms} v4.5 and two absorbed power-law spectral models, as with the {\it Swift} data.  The higher of the two fluxes was used as an upper limit.

The 0.3--10 keV fluxes from all four observatories are displayed in Figure~\ref{lightcurves}, in the form of a long-term light curve spanning $\sim$ two decades ({\it top}), and an expanded version covering the time range of the {\it Swift} monitoring data for clarity ({\it bottom}).  Clearly NGC 5907 ULX is somewhat variable, as the flux is seen to change by factors of $\la 10$ on timescales of months to years.  Multiple epochs of both high and low luminosity are seen, with the {\it Swift\/} data indicative of the luminosity varying between $\sim 0.8$ and $\sim 6 \times 10^{40} \rm ~erg ~s^{-1}$, and a possible gradual decay over most of this range is observed at the start of 2012. 

\subsection{Short-term X-ray timing}

\begin{figure}
\centering
\includegraphics[width=8cm]{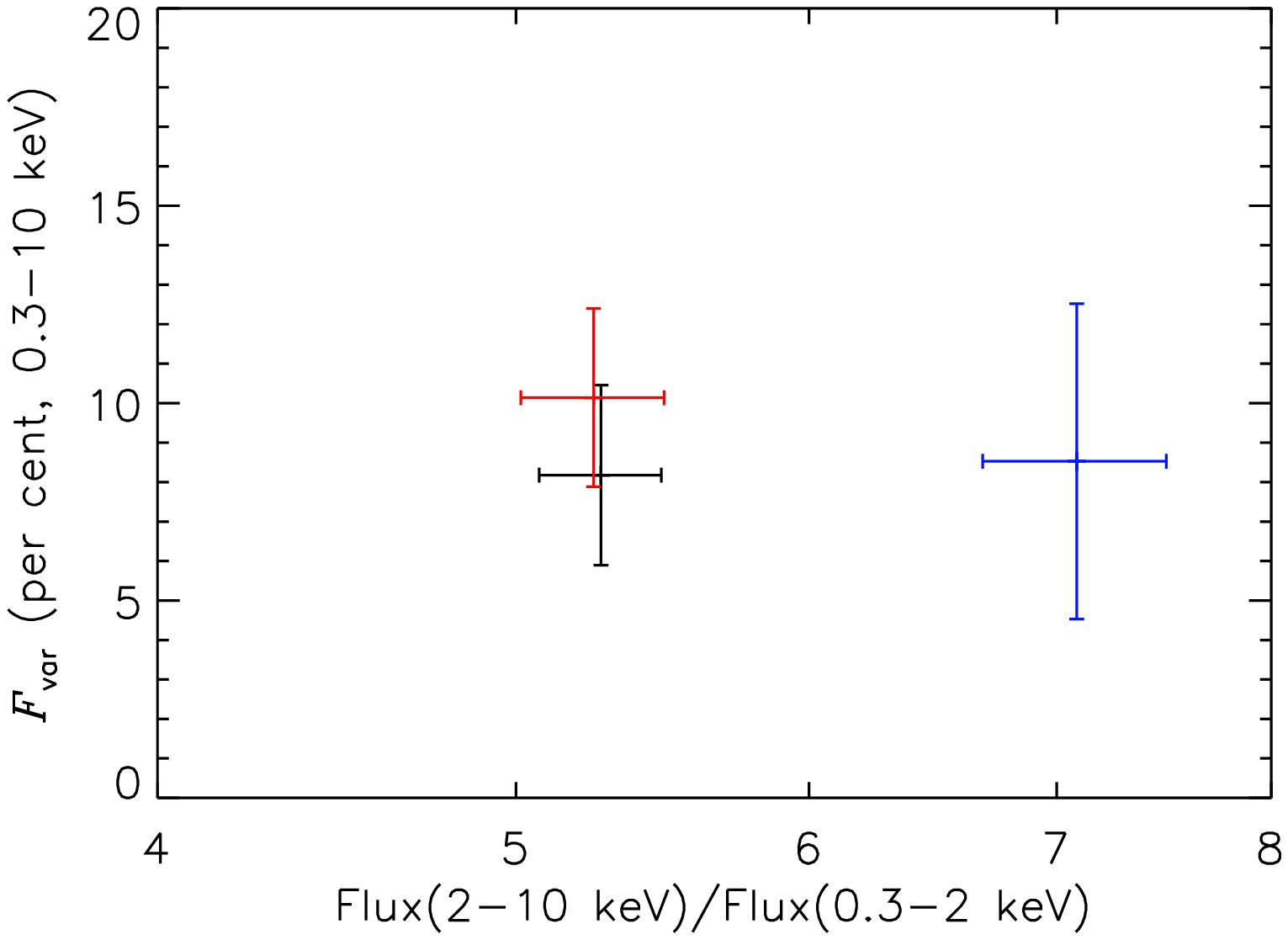}
\includegraphics[width=8cm]{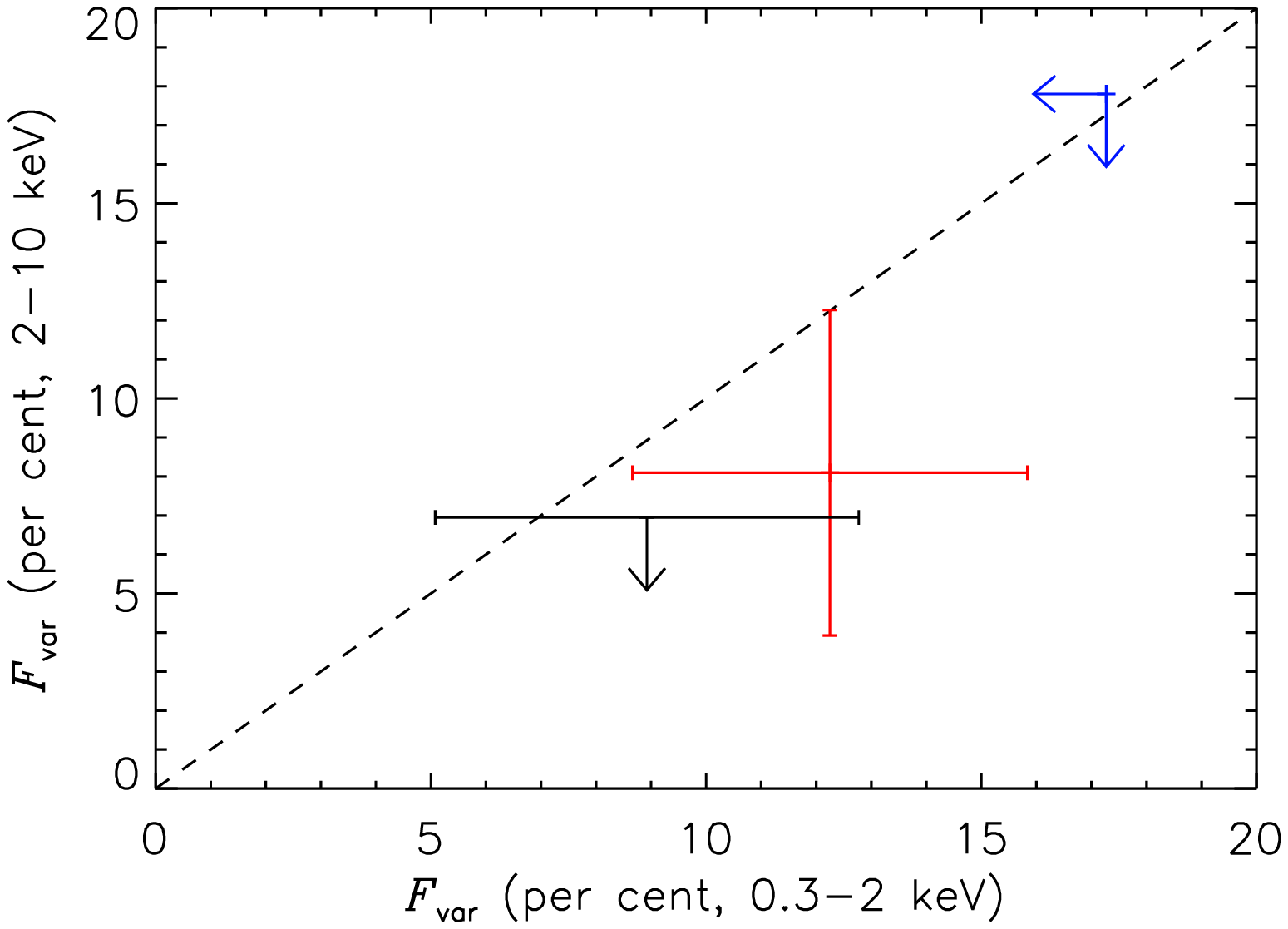}
\caption{The fractional variability of NGC 5907 ULX on a time scale of 200 s, extracted from the {\it XMM-Newton} observations.  {\it Top:\/} fractional variability across the whole energy band (0.3--10 keV) plotted against spectral hardness (see text).  {\it Bottom:\/} soft band (0.3--2 keV) and hard band (2--10 keV) fractional variabilities plotted against each other, where the dashed line shows $F_{\rm var}{\rm (hard)}=F_{\rm var}{\rm (soft)}$.  Observations are identified using colour -- 0145190201 in black, 0145190101 in red and 0673920301 in blue (observation 0673920201 has fewer than twenty continuous 200s temporal bins, so its fractional variability was not calculated).  Errors and upper limits show the $1 \sigma$ uncertainty levels.  The fractional variability of NGC 5907 ULX is consistent (at the $1 \sigma$ level) with remaining constant at the $\sim 9$ per cent level between observations, and in the hard and soft bands.}
\label{fvar}
\end{figure}

An additional diagnostic tool, which can break some of the degeneracy inherent in the X-ray spectral fitting, is fractional variability.  It can be used both to test for consistency with assumed accretion modes, or to confirm the identification of multiple spectral components by their differing variability properties.  Three of the {\it XMM-Newton} observations (0145190201, 0145190101 and 0673920301) contained sufficient good time (a minimum of 20 temporal bins), and were at high enough count rates (resulting in at least 20 counts per temporal bin) to allow us to extract their fractional variability.  We did this with 200 s temporal binning, across all of the available good time, using the definition of fractional variability given by \cite{vaughan_etal_2003}, and excluding bins that contained less than the full 200 s exposure.  Firstly we extracted the broad band (0.3--10 keV) fractional variability from the light curve of each observation to test for changes in timing properties with spectral hardness (Figure~\ref{fvar}, {\it top}), which we define here as the ratio between the fluxes in the 0.3--2 and 2--10\,keV bands.  These fluxes are taken from the disc plus Comptonisation model, and corrected for the known foreground absorption (further corrections are highly uncertain, given the model-dependence of the measured absorption column). The broad band variability over the time scales probed is consistent with being unchanged between observations, despite the changes in spectral hardness.  As a further step we extracted the fractional variability from the soft (0.3--2 keV) and hard (2--10 keV) {\it XMM-Newton} light curves, to test for different behaviour in the soft and hard spectral components (Figure~\ref{fvar}, {\it bottom}).  However, there was no strong evidence of this, with the fractional variability inconsistent between the energy bands at no more than the $1\sigma$ level, although this was not entirely surprising -- the data quality was such that fractional variability was rather poorly constrained when split between two energy bands, and so in three cases it was only possible to place upper limits on it.

\section{Discussion}

We have presented an analysis of the various recent observations of NGC 5907 ULX, examining both its spectral and temporal behaviour as observed by {\it Swift\/}, {\it XMM-Newton\/} and {\it Chandra\/}.  Here, we will attempt to interpret these observations in terms of the underlying physics of the source, primarily using the behaviours seen in other ULXs as a framework for our interpretation.  However, we should start by summarising what was known before we obtained our recent observations.  NGC 5907 ULX was included in the extreme ULX sample of \cite{sutton_etal_2012}, where it was deduced that these rare ULXs possessed observational characteristics that were consistent with IMBHs accreting below the Eddington limit.  However, NGC 5907 ULX stood out from the other objects in that sample due to the identification of a high energy break in its {\it XMM-Newton\/} spectra.  This is not a feature of the known sub-Eddington accretion states, rather it is a key diagnostic of the ultraluminous state, indicative of super-Eddington accretion onto stellar-mass black holes (\citealt{roberts_2007}; \citealt{gladstone_etal_2009}), although in the case of NGC 5907 ULX the extreme luminosities reached may still require a massive stellar black hole (up to 100 $M_{\odot}$, as suggested by e.g. \citealt{zampieri_and_roberts_2009}) as its central engine.

The relatively high absorption column ($\ga 5 \times 10^{21} \rm ~cm^{-2}$ in almost all models) to NGC 5907 ULX creates problems in constraining and interpreting its X-ray spectrum, as it leads to a deficit of counts below 1 keV and so poor statistics at soft energies, even in the high quality {\it XMM-Newton} X-ray spectra (Figure~\ref{spec_plot}).  This dearth of soft counts may also have contributed to the non-detection by {\it ROSAT\/}.  The practical effect of the high column is that it reduces the effective bandwidth that we can examine the source spectrum over, and so limits our ability to constrain the soft end of its spectrum.  This is particularly unfortunate for ULXs, many of which are known to have strong soft excesses in their spectra (e.g. \citealt{stobbart_etal_2006}), that are both a signature of ultraluminous state spectra and a key part of understanding the physics of their super-Eddington emission \citep{gladstone_etal_2009}.  In this work this problem is very evident in our inability to statistically differentiate between a pure Comptonisation spectrum, and a multi-colour disc plus Comptonisation model, and to place firm constraints on the cooler component in the latter model.

However, the data were still very revealing.  We detected significant long-term flux variability, with possible peaks and troughs in the inter-observation light curve of NGC 5907 ULX (Figure~\ref{lightcurves}; Table \ref{flux}), with accompanying X-ray spectral variability also evident between the high quality {\it XMM-Newton} and {\it Chandra} detections.  The light curve, although sparsely sampled, seems consistent with the variation seen in other ULXs that {\it Swift\/} has monitored on timescales of months to years (e.g. \citealt{kaaret_and_feng_2009}; \citealt{strohmayer_2009}; although ESO 243-49 HLX-1 shows a different, regular outburst cycle behaviour, \citealt{lasota_etal_2011}).  One obvious difference is that this variation is at higher luminosities than most other ULXs at between $\sim 0.8$ and $6 \times 10^{40} \rm ~erg~s^{-1}$.  Similar variability patterns are also seen in a number of ULXs with either a mixture of {\it Swift\/} and {\it XMM-Newton\/} monitoring data, or campaigns conducted on {\it XMM-Newton\/} and {\it Chandra\/} alone (e.g. \citealt{roberts_etal_2006}; \citealt{feng_and_kaaret_2006b}; \citealt{grise_etal_2010}), but these data offer a crucial extra dimension -- detailed spectral variability data.  
This has led to the realisation that ULXs can show spectral variability that is degenerate with X-ray luminosity, with this having been demonstrated for NGC 1313 X-1 and Ho II X-1 (\citealt{pintore_and_zampieri_2012}; \citealt{kajava_etal_2012}).  NGC 1313 X-2, on the other hand, has spectral parameters that correlate well with X-ray luminosity \citep{pintore_and_zampieri_2012}.

Unfortunately, the clustering of our {\it XMM-Newton\/} and {\it Chandra\/} spectra means we only have effectively two measurements of the source spectrum, hence we are unable to draw conclusions on whether it also displays degenerate spectral variability.  What we can conclude is that in the data we do have NGC 5907 ULX exhibits spectral hardening as its flux decreases (e.g. Table~\ref{pow}).  Such a trend has also been reported for the $L_{\rm X}$ -- $\Gamma$ correlated ULXs of \cite{kajava_and_poutanen_2009} (for the objects NGC 253 X-4, IC 342 X-1, NGC 1313 X-1, Ho II X-1, Ho IX X-1, NGC 5204 X-1 and NGC 5408 X-1) and \cite{feng_and_kaaret_2009} (NGC 1313 X-1, Ho II X-1 and NGC 5204 X-1, but not IC 342 X-1), although the spectral index of NGC 5907 ULX in its low flux state was unusually hard compared to these other ULXs.  Interestingly, a number of the $L_{\rm X}$ -- $\Gamma$ correlated ULXs also showed spectral pivoting, with three reported to have appeared in low/soft states (NGC 1313 X-1, Ho II X-1, NGC 5204 X-1; \citealt{kajava_and_poutanen_2009}; \citealt{kajava_etal_2012}).  Such spectral evolution is strongly suggestive of a two component spectrum, as is indeed seen for all three of these objects (e.g. \citealt{gladstone_etal_2009}).  All three are in the ultraluminous state, and a detailed inspection of their spectra indicates the pivoting originates in a change in the balance between the soft and hard spectral components (Sutton et al., in prep.).  However, no such pivoting is seen in NGC 5907 ULX -- it remains hard throughout all the observations.  Indeed, an inspection of Table~\ref{spec_par} shows there is no strong statistical necessity for a two-component spectrum for either the high or low flux groups.  This is supported by the absence of a significant difference between the fractional variability in the soft and hard bands in any observation.  Together, they argue that this ULX is dominated by a single spectral component over the observed energy range.

Accretion disc models supplied a statistically acceptable single-component solution to both groups of spectra, with the high flux spectra preferring a non-standard radial emission profile ($p \sim 0.6$) consistent with some degree of advection.  As the normalisation of both the multi-colour disc blackbody model and the $p$-free disc model provide a measure of the inner disc radius, we can use equation 8 of \cite{makishima_etal_2000} to obtain an `X-ray estimated' mass of the black hole in this ULX.  Taking the upper limit on the high flux group from the $p$-free model (this is consistent with both low flux models, and a superior fit to the standard disc for the high flux data), we find that $M_{\rm BH} \alpha \sqrt{\rm cos(\theta)} < 5 M_{\odot}$, where $M_{\rm BH}$ is the black hole mass, $\alpha$ is a positive parameter that equals unity for a Schwarzschild black hole, and is less than unity for a Kerr (i.e. spinning) black hole, and $\theta$ is the inclination angle of the accretion disc to our line-of-sight.  Clearly this is a remarkably low mass estimate for a ULX that exceeds $5 \times 10^{40} \rm~erg~s^{-1}$ in luminosity, although we note that this limit will rise in the case of either the accretion disc not being face-on, or the black hole having significant spin.  The latter may in part be a solution to the anomalously high disc temperatures we observe ($\sim 3$ keV), as the disc would move deeper into the gravitational potential for a Kerr metric and hence heat up \citep{makishima_etal_2000}, although an additional contribution to the high temperatures may also come from a large colour correction factor due to the hot underlying accretion disc \citep{kajava_etal_2012}.  However, even if the high disc temperature can be explained, two problems remain for the disc interpretation.  Firstly, the temperature increases (very significantly, for a standard disc) as the flux diminishes; this is not the expected behaviour of an accretion disc, that classically follows an $L \propto T^4$ relationship \citep*{done_etal_2007}.  Secondly, we do not expect to see variability from an accretion disc on anything other than the longest timescales \citep{wilkinson_and_uttley_2009}; yet we see $\sim 9$ per cent fractional variability across all observations that are well constrained for NGC 5907 ULX.  Thus, we conclude a disc interpretation is unlikely for this object.

We therefore return to the Comptonisation models.  An absorbed Comptonisation model provides a good single-component fit to both of the sets of grouped spectra.  The fit parameters are similar to those seen in the coronae of other ULXs (e.g. \citealt{gladstone_etal_2009}), being both cool and optically thick.  Additionally, the temperature of the corona in NGC 5907 ULX appears to be cooler at higher luminosity.  Although at first seemingly unphysical, such a single component spectrum may be physically understood in the context of the ultraluminous state model of ULXs.  The key to understanding this lies in the physical interpretation, and the geometry of the two components.  Although \cite{gladstone_etal_2009} originally interpreted the two components rather literally as a thick corona above the inner disc, and the direct emission from the cooler outer disc, more recent work has provided evidence to revise this picture.  In particular, it now seems likely the cool component originates in a massive outflowing funnel-shaped wind, as would be expected from a super-Eddington flow, e.g. \cite{poutanen_etal_2007} (see also \citealt{kajava_and_poutanen_2009}, \citeauthor{middleton_etal_2011a} 2011a).  The warmer component may then still be a thick corona above the central regions of the accretion disc, but alternatively it may be the direct emission from the inner regions of the disc itself, where the high temperatures and accretion rates distort the disc spectrum (e.g. \citeauthor{middleton_etal_2011a} 2011a; \citealt{kajava_etal_2012}).  In this case, if we are viewing NGC 5907 ULX at an inclination close to face on, then we are looking directly down the opening angle of the wind, so observe the central emission with a high degree of geometrical beaming \citep{king_2009}.  It then remains to explain the observed spectral variability, possibly this could be interpreted by the corona cooling as more material is raised above the inner disc, so sharing the coronal power across more particles (cf. \citealt{vierdayanti_etal_2010}; for Ho IX X-1; although see \citealt{kong_etal_2010} for an alternate view of the same data).

There is, however a possible flaw with the above interpretation: we detect $\sim 9$ per cent fractional variability in NGC 5907 ULX.  Because, in the Comptonisation-only model, we are essentially looking at an accretion disc that we expect to be invariant, we then again struggle to explain the variability.  However, variability can be produced in this updated ultraluminous state model, where it is understood as originating from variable obscuration of the stable central emission by clumps of material from the wind intercepting the line-of-sight (\citeauthor{middleton_etal_2011a} 2011a).  But, critically, this can only occur at lines-of-sight close to the wind's edge, which appears at odds with the interpretation from the spectral data that we are viewing NGC 5907 ULX close to face-on.  Here, we speculate that this apparent inconsistency can be circumvented if there is an intrinsic disc-like spectral component from the wind, which we are unable to observe due to its intrinsically cool temperature and the unusually high absorption column in the direction of the source.  As such, we use the results of the two component multi-colour-disc plus Comptonisation model, in an attempt to constrain the limits of a possible contribution from the wind.  As the soft spectral data were of insufficient quality to properly constrain a disc component, we instead used characteristic values for ULX disc-like component temperatures of 0.1 and 0.3 keV.  For the warmer of the two disc-like components, the spectral variability appeared mainly to originate in a far greater diminution of the disc component than the corona as the total flux faded; for the cooler disc the spectral hardening plausibly originated in subtle changes in the corona itself.

In this picture the warmer disc would seem to be the less likely scenario for NGC 5907 ULX, as to have the wind vary significantly more than the inner disc region would require at least a contrived geometry, that could extinguish our view of the outer X-ray emitting regions while retaining most of our view to the inner regions.  However, a cooler disc fit might provide a more physically consistent explanation if we consider that, in order to reach the extreme luminosities seen in this object, an extreme super-Eddington flux is required, even for a MsBH.  At high super-Eddington rates one would expect to see a very massive radiatively-driven wind; as this is effectively a blackbody emitter, its temperature is proportional to surface area, and so will reduce as the material in the wind increases \citep{king_and_pounds_2003}, favouring a very cool component.  Such a wind would not only increase the beaming factor, aiding in reaching the extreme luminosities \citep{king_2009}; but also have a narrower opening angle for the collimated region created by the inner disc and wind, potentially leading to a higher probability of any line-of-sight to the central regions crossing the edge of the wind, thus imprinting variability (\citeauthor{middleton_etal_2011a} 2011a).  This scenario, while speculative, is consistent with the picture emerging from other ULXs, and so we favour it as the best current explanation for the properties of this ULX.

One final issue to deal with is the mass of the black hole powering this system.  Other than the high luminosity, there is no evidence in support of an IMBH being harboured by NGC 5907 ULX.  As we have discussed, its characteristics can be plausibly explained by a highly super-Eddington black hole.  The maximal radiative emission by a super-Eddington system has been estimated as $\approx 20 L_{\rm Edd}$ for a face-on system by \cite{ohsuga_and_mineshige_2011}; hence, given the peak luminosity for NGC 5907 ULX of $\sim 6 \times 10^{40} \rm ~erg~s^{-1}$, we are likely to be dealing with a black hole with mass $> 20 M_{\odot}$, i.e. in the MsBH regime.

\section{Conclusions}

Here we have presented a thorough X-ray analysis of NGC 5907 ULX - a highly absorbed, luminous ULX in an edge-on spiral galaxy.  This study has utilised a combination of {\it XMM-Newton}, {\it Chandra}, {\it Swift} and {\it ROSAT} observations, including recently obtained, proprietary epochs of {\it XMM-Newton} and {\it Chandra} data.  The {\it XMM-Newton} and {\it Chandra} X-ray spectral data were well fitted by a variety of spectral models, consistent with both sub- and super-Eddington solutions.  These included accretion disc spectra, possibly with a degree of advection at the highest X-ray luminosities.  However, low estimates of black hole mass based on the warm inner disc temperatures were particularly difficult to reconcile with the extreme X-ray luminosity, which peaks in excess of $5 \times 10^{40}~{\rm erg~s^{-1}}$.  These inconsistencies are compounded by the failure of the source to follow a classical $L_{\rm X} \propto T^4$ relationship, and the seemingly ubiquitous $\sim 9$ per cent variability in its {\it XMM-Newton} light curves.  Rather, spectra dominated by Comptonised emission from a cool, optically thick medium, provide a good fit to the X-ray spectral data, but again offer no explanation to the variability properties.  We therefore speculate that a substantial radiatively-driven wind could be present, in the form of a cool (0.1\,keV) disc-like spectral component, that is hidden from our view by a combination of its cool temperature and the high column to this source.  In this scenario, an extreme (highly super-Eddington) accretion rate produces a massive, radiatively driven wind that both emits as a cool blackbody, and constricts the opening angle of the central regions of the disc.  This narrower collimation may boost the perceived luminosity of the system, and would also mean any line-of-sight to the central regions could be crossed by the clumpy outflowing wind material, imprinting the observed short-term variability onto the stable emission from the inner regions.  The long-term spectral variations then originate in subtle changes of the physical state of the optically-thick coronal component over time.

The X-ray spectral and timing evidence therefore could indicate that NGC 5907 ULX was observed at highly super-Eddington luminosities.  As such, NGC 5907 ULX would not require an IMBH to produce the observed X-ray luminosities, rather the compact accretor is most likely in the MsBH regime.

\section*{Acknowledgements}

We thank the anonymous referee for their useful comments, that helped to improve this paper.  
ADS gratefully acknowledges funding from the UK Science and Technology Facilities Council in the form of a PhD studentship, SAF for a postdoctoral position, and TPR in the form of a standard grant.  
JCG thanks the Avadh Bhatia Fellowship and Alberta Ingenuity.
SAF is also the recipient of an ARC Postdoctoral Fellowship, funded by grant DP110102889.  ER acknowledges the support of a Science Undergraduate Research Experience scholarship from the University of Leicester.
This work is based on observations obtained with {\it XMM-Newton}, an ESA science mission with instruments and contributions directly funded by ESA Member States and NASA; the {\it Chandra} X-ray observatory; the {\it Swift} Gamma Ray Burst Explorer; and {\it ROSAT}.  It has also included observations made with the NASA/ESA Hubble Space Telescope, and obtained from the Hubble Legacy Archive, which is a collaboration between the Space Telescope Science Institute (STScI/NASA) and the Canadian Astronomy Data Centre (CADC/NRC/CSA).

\bibliography{refs}
\bibliographystyle{mn2e}

\bsp

\label{lastpage}

\end{document}